\documentclass[a4paper,10pt]{article}
\usepackage[utf8]{inputenc}
\usepackage{amsmath}
\usepackage[pdftex]{hyperref}
\usepackage[T1]{fontenc}
\usepackage{lmodern}					
\usepackage{amsbsy}
\usepackage{amssymb}
\usepackage{setspace}
\usepackage{wasysym}
\usepackage{graphicx}
\usepackage{geometry}
\usepackage{mathtools}
\usepackage{titling}
\usepackage{authblk}

\usepackage{appendix}
\usepackage[square, sort, numbers]{natbib}
\makeatletter
\@ifundefined{showcaptionsetup}{}{%
 \PassOptionsToPackage{caption=false}{subfig}}
\usepackage{subfig}

\title{Ramifications of disorder on active particles in one dimension}
\author[1,2]{Ydan Ben Dor}
\author[1]{Eric Woillez}
\author[1]{Yariv Kafri}
\author[3]{Mehran Kardar}
\author[4]{\\Alexandre P. Solon}
\affil[1]{Department of Physics, Technion -- Israel Institute of Technology, Haifa 32000, Israel}
\affil[2]{The Russell Berrie Nanotechnology Institute, Technion -- Israel Institute of Technology, Haifa 32000, Israel}
\affil[3]{Department of Physics, Massachusetts Institute of Technology, Cambridge, Massachusetts 02139, USA}
\affil[4]{Sorbonne Université, CNRS, Laboratoire de Physique Théorique de la Matiére Condensée, LPTMC, F-75005 Paris, France}
\date{}

\begin{document}

\maketitle

\begin{abstract}

  The effects of quenched disorder on a single and many active
  run-and-tumble particles is studied in one dimension. For a single
  particle, we consider both the steady-state distribution and the
  particle's dynamics subject to disorder in three parameters: a
  bounded external potential, the particle's speed, and its tumbling
  rate. We show that in the case of a disordered potential, the
  behavior is like an equilibrium particle diffusing on a random force
  landscape, implying a dynamics that is logarithmically slow in
  time. In the situations of disorder in the speed or tumbling rate,
  we find that the particle generically exhibits diffusive motion,
  although particular choices of the disorder may lead to anomalous
  diffusion. Based on the single-particle results, we find that in a
  system with many interacting particles, disorder in the potential
  leads to strong clustering. We characterize the clustering in two
  different regimes depending on the system size and show that the
  mean cluster size scales with the system size, in contrast to
  non-disordered systems.

\end{abstract}

\section{Introduction}

Self-propelled or active particles consume and dissipate energy in order to move persistently. 
The breaking of time-reversal symmetry by the drive, specially in the vicinity
of external boundaries, leads to a plethora of interesting phenomena, distinct from those  in equilibrium systems. For example, {\it E. Coli.}
bacteria swim in circles near planar surfaces~\cite{frymier1995three}, and the motion of active particles is generally rectified by asymmetric objects~\cite{galajda2007wall,tailleur2009sedimentation}. The latter effect generates currents, which in turn lead to
long-range interactions between objects immersed in an active fluid~\cite{baek2018generic}. Closely related is the fact that, in general, the mechanical pressure exerted on confining boundaries does not follow an equation of
state and, on the contrary, depends on the details of the interactions
with the boundary~\cite{Solon2015NatPhys,junot2017active,ben2018forces} and its
curvature~\cite{Fily2014,Nikolai2016PRL}.

So far, the bulk of studies have focused on the physics of active systems in the absence of disorder, namely in uniform environments. However, natural environments of many active agents,
such as bacteria in the gut or enzymes in the intracellular medium~\cite{muddana2010substrate}, are non-uniform. 
While several recent studies have considered the effects of disorder on
models of flocking~\cite{chepizhko2013optimal,toner2018swarming,
toner2018hydrodynamic,das2018polar}, comparatively
less is known for the simpler case of non-aligning active particles subject to different types of quenched disorder. Such systems are realized
experimentally~\cite{volpe2014brownian} (and of course
numerically~\cite{paoluzzi2014run}) using, for example, optical speckle fields or non-smooth substrates. 

To this end, in this paper we consider run-and-tumble particles (RTPs) in a
one-dimensional (quenched) disordered environment. This model of active particles
has the advantage of allowing for exact calculations since one can write explicit expressions for the steady-state distributions and first passage times of non-interacting
RTPs~\cite{Hanggi1984,Solon2015NatPhys}. We discuss three types of
quenched disorder: in the external potential $V(x)$ which is assumed to be bounded, in the speed of the
particles $v(x)$, and in their tumbling rate $\alpha(x)$. The main results are summarized
in Table~\ref{tab:results}.

Most interestingly, a RTP in a bounded random {\it potential} is akin
to a passive Brownian particle in a random {\it force} field.  This
leads to a strongly localized steady-state probability distribution,
with the location of the maximum of the distribution depending on the
exact shape of the potential throughout the system. Moreover, it
exhibits the so-called Sinai diffusion~\cite{sinai1982limit} (reviewed
in~\cite{bouchaud1990classical}) with logarithmically slow spreading
in time. Interestingly, similar behavior has also been predicted for
molecular motors which are stalled by an external
force~\cite{kafri2004dynamics,harms1997driven}.  We remind the reader
that in stark contrast, a passive particle in a bounded
random potential shows normal diffusion with a steady-state
probability distribution which is uniform on large length-scales.

The strong clustering of the probability distribution also has a striking effect on interacting RTPs in the presence of a disordered potential. It is known that in one dimension RTPs with repulsive interactions form clusters of finite size~\cite{Cates2015MIPS,tailleur2008statistical}. Here we argue using simplified models that the picture is very different. We analyze the problem in two regimes, defined below, which we refer to as weak and strong disorder. In the weak disorder case, we show that the density-density correlation function decays {\it linearly} in space with an amplitude which is linear in the system size. In the strong disorder regime we argue for a power-law distribution of cluster sizes, with the average cluster size scaling as the square root of the system size.

When the disorder enters through the particle's speed, the steady-state probability distribution is generically uniform on large length-scales and the spreading is diffusive in time. An interesting exception occurs when the speed distribution is singular near zero (see Table~\ref{tab:results}). Then, the steady-state distribution is peaked at a specific location, with a height which grows as a power law in the system size. The probability distribution in this case spreads with anomalous time exponents.

Finally, for a disordered tumbling rate, the steady-state probability distribution is flat and the spreading is generically diffusive. Similar to the speed disorder, the spreading is anomalous when the tumbling distribution decays to zero with fat tails for large values of $\alpha$  (see Table~\ref{tab:results}).

The paper is organized as follows: we first consider a single particle
in Section~\ref{sec:single particle}, deriving the steady-state
distributions in Section~\ref{sec:steady-state distributions} and
dynamical properties in Section~\ref{sec:MFPT}. The results of
Section~\ref{sec:single particle} are summarized in Table~\ref{tab:results}. 
Next, we turn to the many-body case, which 
we study using simplified models in Section~\ref{sec:many particles}. The discussion is carried for weak disorder in Section~\ref{sec:many-body-weak-disord} and strong disorder in Section~\ref{sec:many-body-strong-disord}. Finally, we conclude and discuss the expected results of the disorder on higher dimensional systems and other active models in Section~\ref{sec:summary}. 

\begin{table}[]
    \centering
    \begin{tabular}{|c|c|c c|}
        \hline
         Disordered & Steady-state distribution, & \multicolumn{2}{|c|}{Mean-square displacement} \\
         		parameter &  for a given realization of disorder  & \multicolumn{2}{|c|}{} \\
         \hline\hline
		 & &  \multicolumn{2}{|c|}{} \\
         $V(x)$ & \multicolumn{1}{l}{$\frac{1}{1-\left(\frac{\mu}{v}\right)^2 \left(\partial_x V\right)^2} \times$} & \multicolumn{2}{|c|}{$\overline{\left<x^2\right>}\sim \log^4\left(t\right)$} \\
		& \multicolumn{1}{r}{$ \exp\left(-\frac{\alpha\mu}{v^2}\intop^x \textrm{d}x'\,\frac{\left(\partial_{x'} V\right)}{1-\left(\frac{\mu}{v}\right)^2 \left(\partial_{x'} V\right)^2}\right)$} &  \multicolumn{2}{|c|}{} \\      
         & & \multicolumn{2}{|c|}{} \\
         \hline
         & &  \multicolumn{2}{|c|}{} \\
         $v(x)$ & $\frac{1}{v(x)}$ &
          \multicolumn{1}{l}{$p(v)\underset{v\to 0}{\to}0:$} & \multicolumn{1}{l|}{$\overline{\left<x^2\right>}\sim t$} \\
         & &  \multicolumn{2}{|c|}{} \\
         & &
          \multicolumn{1}{l}{$p(v)\underset{v\to 0}{\sim}v^{-\beta},\ 0<\beta<1:$} & \multicolumn{1}{l|}{$\overline{\left<x^2\right>}\sim t^{1-\beta}$} \\ 
          & & \multicolumn{2}{|c|}{}  \\
          \hline
          & &  \multicolumn{2}{|c|}{} \\
         $\alpha(x)$ & const. & \multicolumn{1}{l}{$p(\alpha)\underset{\alpha\to \infty}{\sim}0:$} & \multicolumn{1}{l|}{$\overline{\left<x^2\right>}\sim t$} \\ 
         & &  \multicolumn{2}{|c|}{} \\
         & & \multicolumn{1}{l}{$p(\alpha)\underset{\alpha\to \infty}{\sim}\alpha^{-(1+\mu)},\ 0<\mu<1:$} & \multicolumn{1}{l|}{$\overline{\left<x^2\right>}\sim t^{\frac{2\mu}{1+\mu}}$} \\
         & & \multicolumn{2}{|c|}{} \\
         \hline
    \end{tabular}
   \caption{Summary of results for the single particle problem: For each
      form of disorder, the steady-state distribution is given for
      a specific realization, along with the scaling
      of the variance of the probability density in time. The
      mean-square displacement is averaged both over histories of the
      dynamics, denoted by $\langle\cdot\rangle$, and over all
      realizations of the disorder, denoted by
      $\overline{\,\cdot\,}$.} \label{tab:results}
\end{table}

\section{Single-particle problem}\label{sec:single particle}

We consider a Run-and-Tumble Particle (RTP) in one dimension. The particle moves either to the right or to
the left with a speed $v(x)$ and switches
direction (tumbles) at rate $\alpha(x)/2$. We allow both the speed and
tumbling rate to depend on the position $x$ of the particle. Finally,
the particle experiences an external potential $V(x)$. The probability
density $P_+(x,t)$ ($P_-(x,t)$) to find a right (left)
moving particle at position $x$ at time $t$ is determined by the
Fokker-Planck
equation~\cite{berg1993random,schnitzer1993theory,Tailleur2008PRL}
\begin{align}
    \partial_t P_+(x,t)=&-\partial_x \left[v(x) P_+(x,t)-\mu (\partial_x V) P_+(x,t)\right] -\frac{\alpha(x)}{2}\left[P_+(x,t) - P_-(x,t)\right]\ , \nonumber \\
	\label{eq:RTP FP Model} \partial_t P_-(x,t)=&-\partial_x \left[-v(x) P_-(x,t)-\mu (\partial_x V) P_-(x,t)\right] -\frac{\alpha(x)}{2}\left[P_-(x,t) - P_+(x,t)\right]\;,
\end{align}
with $\mu$ the mobility of the particle. To avoid trivial trapping, we
assume that $v(x)>|\mu\partial_x V|$. This condition can be avoided
if, in addition, the particle is subject to Brownian noise. The noise
can assist the particle in hopping over potentials of arbitrary
slope~\cite{woillez2019activated}. However, as we discuss in
Sec.~\ref{sec:summary}, we do not expect fundamentally new physics in
this case, and we thus consider only the technically simpler noiseless
situation.

We next consider the effect of disorder on the steady-state
distribution in Section~\ref{sec:steady-state distributions} and on
the dynamics in Section~\ref{sec:MFPT}. In each case, we
focus on the three different types of disorder: in the potential, the
speed and the tumbling rate.

\subsection{Steady-state distributions}\label{sec:steady-state distributions}

\begin{figure}
\begin{center}
	\includegraphics[width=.95\textwidth]{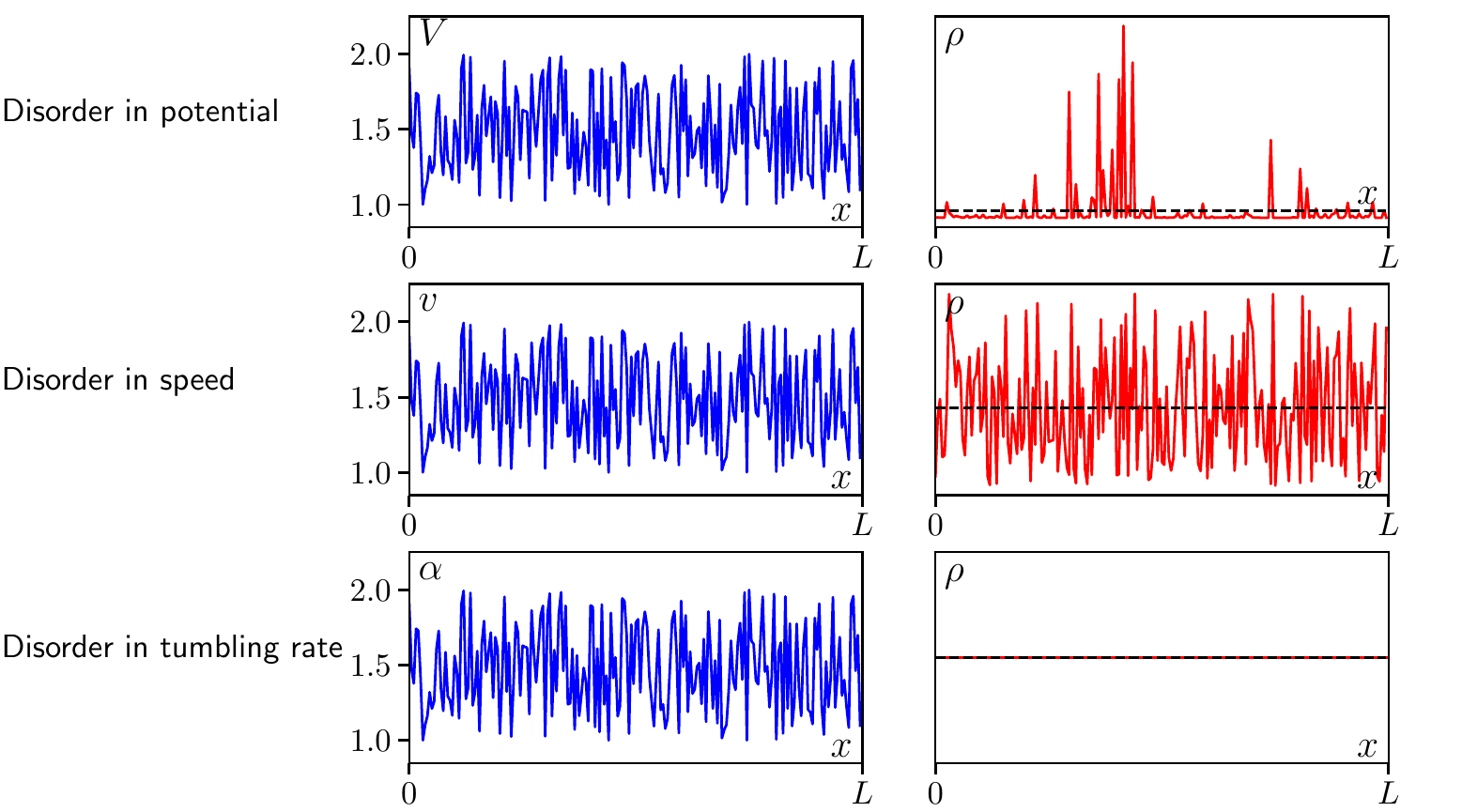} 
	\caption{Steady-state distributions (right column) for the
          disorder profiles shown in the left column (for system size
          $L=200$). Top row: $v=2$, $\alpha=40$; middle and bottom
          rows: $V=0$.}\label{fig:example-ss}
      \end{center}
\end{figure}

For the dynamics of Eq.~\eqref{eq:RTP FP Model}, the steady-state
distribution can be computed
analytically~\cite{Hanggi1984,Solon2015NatPhys}. One starts by
defining the total probability density $\rho(x,t)=P_+(x,t)+P_-(x,t)$
and the polarity $\Delta(x,t)=P_+(x,t)-P_-(x,t)$. Using Eq.~\eqref{eq:RTP
  FP Model}, we have
\begin{align}\label{eq:steady state rho and delta}
	\partial_t \Delta(x,t) = & -\partial_x\left[v(x) \rho(x,t) -\mu (\partial_x V) \Delta(x,t)\right] - \alpha(x)\Delta(x,t)\ ,\nonumber\\
	\partial_t \rho(x,t) = & -\partial_x\left[v(x) \Delta(x,t) -\mu (\partial_x V) \rho(x,t)\right]\ .
\end{align}
Assuming that the system is confined, so that there is no particle
current in the steady state, one finds for a given realization of
$\alpha(x),\,v(x)$ and $V(x)$ that the steady-state 
density $\rho_s(x)$ is given by
\begin{equation}\label{eq:general steady-state distribution}
	\rho_s(x) = \frac{{\cal N} v(x)}{v^2(x)-\mu^2 \left(\partial_x V\right)^2} \exp\left(-\intop_0^x \textrm{d}x'\,\frac{\alpha(x') \mu \left(\partial_{x'} V\right)}{v^2(x')-\mu^2 \left(\partial_{x'} V\right)^2}\right)\ ,
\end{equation}
where ${\cal N}$ is a normalization constant set by $\intop
\textrm{d}x\,\rho_s(x)=1$. Equation~(\ref{eq:general steady-state
  distribution}) allows for variations 
   in the potential $V(x)$, the speed $v(x)$
and the tumbling rate $\alpha(x)$. Figure~\ref{fig:example-ss} shows the
resulting density if each component acts individually.
It is easy to see from the figures that different types of disorder lead to very different behaviors: disorder in the potential has the strong effect of localizing the probability distribution to several locations in space; disorder in the
velocity leads to a local modulation of the density which depends on the local velocity;
while disorder in the tumbling rate leads to a flat steady-state distribution. In the following, we discuss these three cases in more detail.

\subsubsection{Disordered potential}
\label{sec:potential disorder steady state}

We first consider a disordered potential $V(x)$ while maintaining constant speed  $v(x)=v$ and tumbling rate
$\alpha(x)=\alpha$. As stated above, we choose the potential such that $|\mu\partial_x V|<v$ so that the particle can
cross any potential barrier. Let
us first rewrite Eq.~(\ref{eq:general steady-state distribution})
in terms of $\tilde V(x)\equiv \mu V(x)/v$ (so that the force
$\partial_{x}\tilde V$ is dimensionless), and the
inverse run length $g\equiv \alpha/v$, as
\begin{equation}\label{eq:steady state potential}
	\rho_{s}\left(x\right)=  \frac{{\mathcal{N}_p} }{1-\left(\partial_{x}\tilde V\right)^{2}}\exp\left(-g \intop_{0}^{x}\textrm{d}y\,\frac{\partial_{y}\tilde V}{1-\left(\partial_{y}\tilde V\right)^{2}}\right)
\end{equation}
with $\mathcal{N}_p$ ensuring normalization. We further assume that $V(x)$ has only short-range correlations and
denote its correlation length by $\xi$. We can then approximate the
integral in Eq.~(\ref{eq:steady state potential}) as a sum of
independent identically distributed (i.i.d.) random variables
\begin{equation}\label{eq:sum-rdvar}
	\intop_{0}^{x}\textrm{d}y\,\frac{\partial_{y}\tilde V}{1-\left(\partial_{y}\tilde V\right)^{2}}\approx \sum_{i=1}^{\lfloor x/\xi\rfloor}\eta_i\ .
\end{equation}
The full justification of the above approximation can be found e.g. in \cite{gardiner1986handbook}. To discuss the resulting stationary distribution, it is instructive to rewrite Eq.~(\ref{eq:steady state
  potential}) as an equilibrium distribution
$\rho_s(x)\propto e^{-U(x)}$ with the quasi-potential
\begin{equation}\label{eq:pot-U}
	U(x)\approx \ln\left[1-\left(\partial_{x}\tilde V\right)^{2}\right]+g\sum_{i=1}^{\lfloor x/\xi\rfloor}\eta_i\ .
\end{equation}
Since the sum in Eq.~(\ref{eq:pot-U}) is over i.i.d. random variables,
the quasi-potential scales as $U(L)\propto\sqrt{L/\xi}$ by
virtue of the central limit theorem, given that the system is large enough $L\gg \xi$. For such systems, the quasi potential is dominated by 
the sum over the random variables and the logarithmic local term is negligible. At the level of the
steady-state distribution, the  behavior is thus identical to that of
a passive particle in a {\it random force field}, with $g\eta_i/\xi$ as the
random force. For any realization of such a  {\it random force}, 
as in Fig.~\ref{fig:example-ss} (top row),
the particle is strongly localized to the minima of the quasi-potential, 
which are of order $\sqrt{L}$.

The difference between the passive and active cases is therefore dramatic:
an active particle in a bounded random potential is equivalent to a passive
particle in a random force field. By contrast, a passive particle in a
bounded random potential $V(x)$  does not show
localization. The difference due to the activity can be understood
intuitively as follows: since RTPs break time reversal symmetry, the
particle exerts a net force on a potential lacking inversion symmetry (this effect was used
experimentally to propel asymmetric objects through a bacterial
bath~\cite{DiLeonardo2010PNAS,Sokolov2010PNAS}). Conversely, the
external potential exerts a net force on the RTP so that a random
potential will effectively lead to a net random force~\cite{Nikolai2016PRL}. For example, if we choose a particularly simple realization of the potential, where ratchet potentials of opposing directionality are drawn with equal probability, as illustrated in Fig. \ref{fig:ratchet potential}, the orientation of each ratchet biases the active particle in a particular direction acting as a local force. We note that similar physics applies for other ratchet-like systems~\cite{kafri2004dynamics,harms1997driven}.

The model of a random walker subject to random forcing is known as
the Sinai diffusion problem, and has been studied extensively in the
past with many applications
\cite{sinai1982limit,derrida1983velocity,bouchaud1990classical,le1999random}. In
Section~\ref{sec:MFPT potential}  we show that the equivalence between active particles in a random potential and passive particles in random force field also extends to
the particle's dynamics.

\subsubsection{Disordered speed}\label{sec:velocity steady state}

We now consider systems where the disorder enters only through a
space-dependent speed $v(x)$, while the tumbling rate $\alpha$ is
constant and there is no external potential. We take $v(x)$ to be a
random field taking values in the range $0 < v(x) <\infty$, with only
short-range correlations and a probability distribution
independent of space\footnote{Note that for the following results to hold, the correlations in $v(x)$ need not be short ranged. This demand will play a important role only for the dynamics.}. For a particular realization of the disorder, Eq.~(\ref{eq:general
  steady-state distribution}) becomes
\begin{equation}\label{eq:steady state speed}
	\rho_s(x) = \frac{\mathcal{N}_v}{v(x)}\ ,
\end{equation}
where $\mathcal{N}_v$ insures the normalization of the
distribution. We note that such systems have been extensively studied in the 
past~\cite{schnitzer1993theory,Tailleur2008PRL} and used experimentally to
draw patterns using photokinetic
bacteria~\cite{frangipane2018dynamic,arlt2019dynamics}.

Evidently, the active particle density is enhanced in places
where the speed of the particles is smaller. However, in contrast with
the case of a random potential discussed above, the steady-state
distribution of the random speed model is a {\it local} function of $v(x)$
(whereas Eq.~\eqref{eq:steady state potential} shows that the density  is a non-local functional of the potential). The resulting
distribution~\eqref{eq:steady state speed}, while non-uniform, is therefore
not strongly localized in general. The particles show similar
statistics as an equilibrium system subject to a random potential, $U(x)=-\ln v(x)$,
with a finite variance.

An exception occurs when $v(x)$ takes values arbitrarily close to zero.
In particular, consider a probability distribution for $v$ that vanishes
as $p(v)\underset{v\to0}{\sim}v^{-\beta}$ close to $0$.
When $0<\beta<1$ the mean of
$v^{-1}$ diverges. Then, using standard arguments from L\'evy
statistics, the largest value of $v^{-1}$ in a system of size $L$
scales\footnote{This can be seen by evaluating
  $\intop_0^{v^*}\frac{dv}{v^\beta}\sim\frac{1}{L}$, with $v^*$ the largest value of $v$ that is observed.} as
$L^{\frac{1}{1-\beta}}$. The particle becomes localized at the position of the maximal value of $v^{-1}$,
with a probability scaling in the same way.
Note that the non-uniformity of $\rho_s(x)$  is due to the singular distribution of velocities, and is not a cumulative effect of a non-local dependence  on $v(x)$.

\subsubsection{Disordered tumbling rate}
\label{sec:tumbling rate steady state}

Finally, we consider quenched disorder only in the
tumbling rate. We take $\alpha(x)$ to be a random field which takes
values in the range $0 < \alpha(x) <\infty$ with short-range
correlations, independent of space. Then, using Eq.~\eqref{eq:general
  steady-state distribution} with $\partial_x V=0$, $v(x)=v$ we
immediately get
\begin{equation}\label{eq:Random tumbling rate steady state}
	\rho_s(x) = \text{constant}=\frac{1}{L}.
\end{equation}
Disorder in $\alpha$ leaves the steady-state distribution flat.

\subsection{Dynamics through the mean first passage time}\label{sec:MFPT}

\begin{figure}
  \begin{center}
    \includegraphics[width=.7\textwidth]{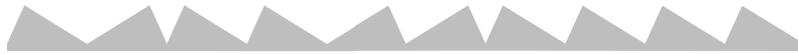} 
    \caption{A realization of the ratchet potential used in the numerics. The two possible orientations of each ratchet are chosen with equal probability.}\label{fig:ratchet potential}
  \end{center}
\end{figure}

We now turn to analyze the effects of disorder on the dynamics of  active particles. 
This is most easily accomplished by considering the typical, disorder-averaged, mean first passage time (MFPT) for a particle to travel a distance $L$ in either direction starting at an arbitrary point. As we show below, the insights gained by studying the steady-state distribution can also be extended to the dynamics. Specifically, disorder in the potential leads to behavior similar to a random walker on a random \emph{forcing} energy landscape. We show in Section \ref{sec:MFPT potential} that the typical MFPT grows {\it exponentially} with $L^{\frac{1}{2}}$. This leads to an ultra-slow diffusive dynamics of the particle with  a typical mean-square displacement growing in time as $\ln^4(t)$. 

In contrast, disorder in the speed or tumbling rate generally leads to a behavior similar to equilibrium dynamics of a random walker in a bounded random potential. The  MFPT averaged over the disorder follows a standard diffusion law, and grows as $L^2$. As discussed in Section~\ref{sec:velocity steady state}, an exception occurs for a disordered speed distribution in which the mean inverse speed, $\overline{v^{-1}}$, over disorder realizations diverges, and for disordered tumbling rates when the mean tumbling rate $\overline{\alpha}$ diverges. 
These cases exhibit anomalous  behavior, with the disorder-averaged mean-square displacement scaling with a non-trivial power of time.

\subsubsection{Disordered potential}\label{sec:MFPT potential}

We now evaluate the typical MFPT for an active run-and-tumble particle in a random potential with uniform  speed and tumbling rate. We consider a particle starting at an arbitrary point, taken to be $x=0$, and exiting either at $x=L$ or $x=-L$.
Similar to the calculation for the steady-state distribution, we define the quasi-potential 
\begin{equation}
	W(x)=\frac{\mu \alpha}{v^2}\intop_{-L}^x \textrm{d}y\,\frac{\partial_y V}{1-\left(\frac{\mu}{v}\right)^2\left(\partial_y V\right)^2}.
\end{equation}
Using the derivation presented in Appendix~\ref{app:computation of MFPT random potential}, one finds that in the large $L$ limit the MFPT, up to exponentially smaller corrections in $L$,  is given by
\begin{equation}\label{eq:MFPT dominating term potential}
	 \left<\tau\right> \underset{L\rightarrow\infty}{\sim} 
	 \frac{\intop_{-L}^0 \textrm{d}x\,\intop_0^L \textrm{d}y\,\intop_{x}^y \textrm{d}z\, Q(x)Q(y)Q(z) \exp\left[ -W(x) - W(y) + W(z) \right]}
	 {\intop_{-L}^{L}  \textrm{d}u\, Q(u)\exp\left[-W(u)\right]}\ .
\end{equation}
Here, $\langle \tau \rangle$ is the MFPT for a given realization of disorder, with the angular brackets denoting an average over histories, 
$Q(x)$ is a non-vanishing function whose expression is given in Appendix~\ref{app:computation of MFPT random potential} and, as will become clear, does not influence the leading order behavior.
 
Since the potential $V(x)$ is assumed to have short-range correlations, $W(x)$ is a sum of i.i.d. random variables. Note that since the integrand 
in $W(x)$ is anti-symmetric with respect to inverting $\partial_y V(x)$ to $- \partial_y V(x)$, the average of $W(x)$ over the realizations of the disorder vanishes. The central limit theorem then gives the behavior of $W(x)$ in the large $L$ limit: the distribution of $W(x)$ converges asymptotically to a Gaussian distribution, whose variance scales as $\sqrt{L/\xi}$, with $\xi$ being the correlation length of $V(x)$, as before. In the large $L$ limit, the MFPT is therefore dominated by the exponential term and can be evaluated using a saddle-point approximation. One then finds
\begin{equation}\label{eq:MFPT SPA}
	\ln\langle\tau\rangle \underset{L\rightarrow\infty}{\sim} -\min_{x\in[-L,0],y\in[0,L],\,z\in[x,y]} \left[W(x)+W(y)-W(z)\right] +\min_{u\in[-L,L]} \left[ W(u)\right] \ .
\end{equation}
Even if the MFPT~\eqref{eq:MFPT SPA} has a non-trivial dependance on the potential, it is clear that its asymptotic behavior in the large $L$ limit is of the order of the largest difference of the effective potential $W(x)$ in the $[-L,L]$ interval\footnote{The result is easier to interpret if one computes the simplified MFPT, obtained by imposing reflecting boundary condition at the origin. Then a calculation similar to the one presented here shows that the logarithm of the MFPT is dominated by the largest difference in the effective potential $W(x)$. As the scaling with $L$ is the same, the discussion that follows is identical.}.

Note that for a given realization of the disorder, the MFPT is controlled by an exponentially large quantity in the potential difference. Therefore, there is a difference between the average MFPT and the typical one. The latter is of interest and is encoded in the disorder average of $\ln\langle\tau\rangle$ given in Eq.~\eqref{eq:MFPT SPA}. 
This gives
\begin{equation}\label{eq:MFPT Sinai scaling}
	\overline{\ln\left<\tau\right>} \underset{L\rightarrow\infty}{\sim} A \sqrt{L}\ ,
\end{equation}
where $A$ is a constant that depends on the details of the potential $W(x)$.  

\begin{figure}
\begin{center}
	\includegraphics[width=.6\textwidth]{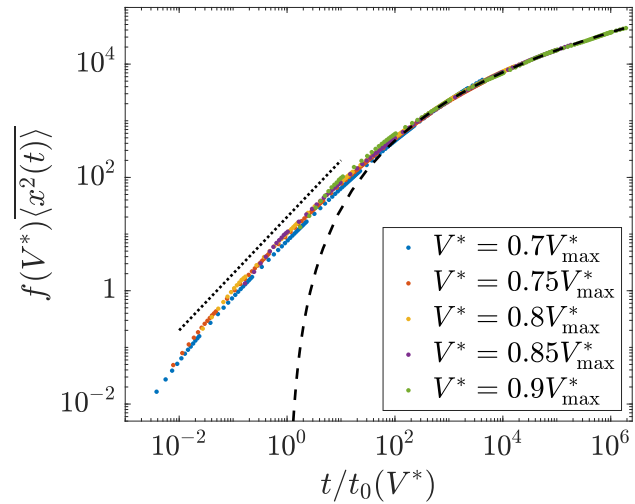} 
	\caption{
	Dynamics of active particles on random ratchet potentials of varying strength. Different plots correspond to ratchets which differ only by their maximal height, $V^*$, with a slope ratio of 1:4. Heights are given relative to the maximal ratchet height $V^*_{\rm max}$ which imposes a slope $|\mu\partial_xV|=v$. Such slope prevents the particle from moving.
	 For each $V^*$, 100 RTPs are simulated on a potential of $10^4$ ratchets for $10^{7}$ time steps, with unit mobility and speed. The time and mean-square displacement are rescaled by constants so that the data collapse for long times. The dashed line marks the theoretical prediction of Eq.~\eqref{eq:spreading} $\overline{\langle x^2(t)\rangle}\propto\ln^4(t)$, showing good agreement on long times with the numerical data. The dotted line shows the diffusive scaling $\overline{\langle x^2(t)\rangle}\propto t$, valid for short times. For stronger $V^*$, the numerical curves approach the disorder dominated regime earlier.}\label{fig:Sinai single}
	\end{center}
\end{figure}

The above result indicates that the mean-square displacement of an active particle on a random potential behaves as 
\begin{equation}\label{eq:spreading}
	\overline{\langle x^2(t)\rangle}\underset{t\rightarrow\infty}{\propto} \ln^4(t)	
\end{equation}
where $x(t)$ is the displacement at time $t$.
This indicates that the dynamics of the active particle on a one dimensional  random potential energy landscape is an ultra-slow Sinai diffusion. Indeed, at the exponential level, the MFPT of the two models is identical.
We verify this prediction numerically in Fig.~\ref{fig:Sinai single} for the random ratchet model illustrated in 
Fig.~\ref{fig:ratchet potential}: the disorder-averaged mean-square displacement as a function of time for active RTP on a disordered random ratchet potential agrees with Eq.~\eqref{eq:spreading} in the long-time limit. 

It is interesting to ask when the effects of a weak random potential
become important.  To study this question we appeal to the equilibrium
random forcing analogy. In this case, a natural length
scale~\cite{bouchaud1990classical} for the crossover is given by
$\ell^* \simeq D^2/\sigma^2$ with $D$ the diffusivity of the particle
and $\sigma^2$ the variance of the random force. Namely,
$\overline{f(x)f(x')}=\sigma^2 \delta(x-x')$, with $f(x)$ the force at
position $x$. In the active case $D=v^2/\alpha$, while $\sigma^2$
depends on the details of the potential distribution. The value of
$\sigma^2$, which depends in a non-trivial way on the parameters of
the model and cannot be easily evaluated, is a measure of the strength
of the ratchet effect for a given disorder distribution. Therefore, as
expected, the stronger ratchets lead to shorter crossover lengths (see
Fig. \ref{fig:Sinai single}). Note that the same length scale can be
obtained from Eq.~(\ref{eq:steady state potential}) by considering the
quasi-potential. In analogy with a Boltzmann weight, the temperature
scale $T$ is set by the ratio $\sigma\sqrt{L}/D$, where
$D=v^2/\alpha$. The low temperature regime then corresponds to the
strong disorder limit $L\gg\ell^*$ while the high temperature limit
corresponds to systems where $L\ll\ell^*$ and the density profile is
to leading order uniform.

\subsubsection{Speed disorder}
\label{sec:velocity MFPT}

We now turn to  the dynamics of  active particles in the presence of a spatially varying speed. The difference in steady-state distributions between this case and that of random potentials was already emphasized in Sec.~\ref{sec:velocity steady state}. To proceed, we note, using the results from Appendix~\ref{app:computation of MFPT random speed and tumbling rate}, that when the disorder enters through the speed, the MFPT is asymptotically given by
\begin{equation}\label{eq:MFPT random speed}
	\left<\tau\right> \underset{L\rightarrow\infty}{\sim} \frac{\alpha\intop_{-L}^0\textrm{d}x\,\,\intop_{-L}^L\textrm{d}y\,\intop_x^y\textrm{d}z\,v^{-1}(x)v^{-1}(y)v^{-1}(z)}{\intop_{-L}^L\textrm{d}z\,v^{-1}(w)}\ .
\end{equation}
This expression will be analyzed in two distinct cases. In the first, the speed probability density vanishes near $v=0$, while in the second the probability density diverges near $v=0$.

\begin{figure}
\begin{center}
	\includegraphics[width=.5\textwidth]{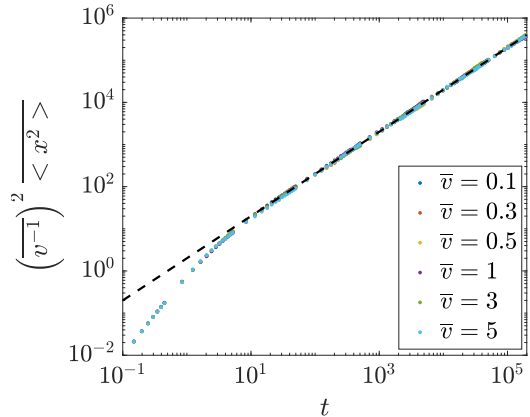} 
	\caption{Dynamics of active particles with varying speed, randomly drawn at
	each point from a uniform distribution in the interval $[0.9\overline{v},1.1\overline{v}]$. The plots are for 100 particles and $10^8$ time steps, 1000 realizations of disorder, 6 mean speeds and a unit tumbling rate. The disorder-averaged variance of the probability distribution, rescaled according to Eq.~\eqref{eq:tau speed}, is plotted as a function of time. The dashed line is fitted to diffusive motion, $\overline{\langle x^2(t) \rangle} \sim t$.}\label{fig:random speed}
	\end{center}
\end{figure}

\noindent {\bf Case I: vanishing probability density near $\mathbf{v=0}$.}
In this case, the dynamics is diffusive.
The denominator of Eq.~\eqref{eq:MFPT random speed} in the large $L$ limit is well approximated, due to the law of large numbers, by $2L \overline{v^{-1}}$. Then, the disorder average of $\langle\tau\rangle$ can readily be evaluated. The leading order term in the numerator is given by the product of the disorder averages of the inverse speed, scaling as $L^3$, with corrections  scaling as $L^2$. To see this, one defines $v^{-1}=\overline{v^{-1}}+\delta v^{-1}$ and assumes short-range correlations of the deviations $\delta v^{-1}$ in space. This leads to diffusive behavior,
\begin{equation}\label{eq:tau speed}
	\overline{\left<\tau\right>} \underset{L\rightarrow\infty}{\sim} \frac{\alpha}{2} \left(\overline{v^{-1}}\right)^2L^2\ ,
\end{equation}
as verified numerically in Fig.~\ref{fig:random speed}.

\begin{figure}
\begin{center}
	\includegraphics[width=.5\textwidth]{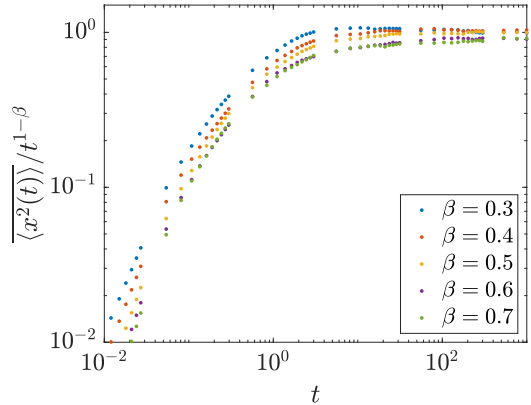} 
	\caption{The disorder-averaged spreading for randomly distributed speed (case II), rescaled with time as suggested by Eq.~\eqref{eq:MFPT random speed divergent scaling}. The dynamics of active particles is simulated with  speeds varying in space, drawn from the distribution $p(v)\sim v^{-\beta}$ in the interval $[0,1]$, for 100 particles  $10^5$ time steps,  1000 realizations of disorder, and unit tumbling rate.}\label{fig:random speed divergent}
	\end{center}
\end{figure}

\noindent{\bf Case II: diverging probability density near $\mathbf{v=0}$.}
Here we assume that the probability density near $v=0$ takes the form $p(v)\sim v^{-\beta}$, with $0<\beta<1$. It is well known that slow bonds can lead to anomalous diffusive behavior~\cite{alexander1981excitation}, and a similar phenomenon also occurs here. To evaluate the MFPT~\eqref{eq:MFPT random speed}, we note that the dependence of the denominator on $L$ can be obtained using standard properties of L\'evy distributions~\cite{bouchaud1990classical}. The integral of the denominator of Eq.~\eqref{eq:MFPT random speed} is dominated by the largest value of $v^{-1}$ on the interval of length $L$, denoted by $(v^*)^{-1}$, which scales as $(v^*)^{-1}\approx L^{\frac{1}{1-\beta}}$. Similarly, the numerator is dominated by the largest contribution to each of the integrals, scaling as $L^{\frac{3}{1-\beta}}$. The resulting MFPT is then given by
\begin{equation}\label{eq:MFPT random speed divergent scaling}
	\overline{\left<\tau\right>} \underset{L\rightarrow\infty}{\propto} \frac{\alpha}{2} L^{\frac{2}{1-\beta}}\ .
\end{equation}
This anomalous diffusion is verified numerically in Fig.~\ref{fig:random speed divergent}\footnote{For the largest value approximation to be valid, the extreme value must
exceed the average contribution from the integral, which is proportional to $L$. The corrections to the scaling of the numerator as a function of $L$ behave as $L^{\frac{2-\beta}{1-\beta}}$. Therefore, in numerics, one has to look at system sizes such that $L^{\frac{2}{1-\beta}}\gg L^{\frac{2-\beta}{1-\beta}}$ or $L^{\frac{\beta}{1-\beta}}\gg 1$, which becomes difficult as $\beta\to 0$.}. Finally, we note that the results suggests that when $\beta=0$, one should expect $\overline{\langle \tau\rangle}\propto L^2/\log(L)$.

\subsubsection{Tumbling rate disorder}\label{sec:tumbling rate MFPT}

Finally, let us consider the dynamics of the active particle in the
presence of a tumbling rate that varies in space. This was shown in
Section~\ref{sec:tumbling rate steady state} to have no effect on the
steady-state distribution, leading to a flat density profile.

Using the results of Appendix~\ref{app:computation of MFPT random speed and tumbling rate}, to leading order in $L$ the MFPT is given by 
\begin{equation}\label{eq:MFPT random tumbling rate}
	\langle\tau\rangle \underset{L\to\infty}{\sim} \frac{1}{v^2}\frac{\intop_{-L}^0\textrm{d}x\,\intop_{0}^{L}\textrm{d}y\,(y-x)\alpha(x)\alpha(y)}{\intop_{-L}^L\textrm{d}z\,\alpha(w)}\ .
\end{equation}
As in the case of the speed disorder, we distinguish here between two limits.

\begin{figure}
\begin{center}
	\includegraphics[width=.5\textwidth]{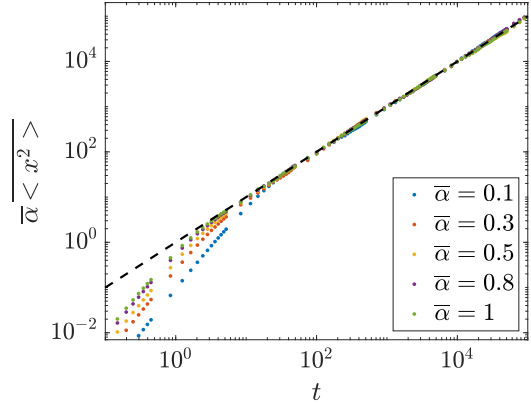} 
	\caption{Dynamics of active particles with a random tumbling rate varying in space (case I). The tumbling rate is drawn from a uniform distribution in $[0,2\overline{\alpha}]$, for 100 particles, $10^8$ time steps, 1000 realizations of disorder, 5 mean tumbling rates, and a unit speed. The disorder-averaged variance of the probability distribution, rescaled according to Eq.~\eqref{eq:tau tumbling rate}, is plotted as a function of time. The curves are fitted with diffusive dynamics, i.e. $\overline{\langle x^2(t) \rangle} \sim t$.}\label{fig:random tumbling rate}
	\end{center}
\end{figure}

\noindent{\bf Case I: probability density with a finite mean.}
Here, we can use an integration by part to transform Eq. \eqref{eq:MFPT random tumbling rate} into
\begin{equation}\label{eq:MFPT transformed}
	\langle\tau\rangle \underset{L\to\infty}{\sim} \frac{1}{v^2}\frac{\intop_{-L}^0\textrm{d}x\,\intop_{0}^{L}\textrm{d}y\,\left(\alpha(x)\intop_{y}^{L}\textrm{d}z\,\alpha(z)+\alpha(y)\intop_{-L}^{x}\textrm{d}z\,\alpha(z)\right)}{\intop_{-L}^L\textrm{d}z\,\alpha(w)}\ .
\end{equation}
Then using the law of large numbers and arguments almost identicals to those leading to Eq. (\ref{eq:tau speed}), we find that the MFPT is given by
\begin{equation}\label{eq:tau tumbling rate}
	\overline{\left<\tau\right>} \underset{L\rightarrow\infty}{\sim} \frac{\overline{\alpha}}{2v^2}L^2\ .
\end{equation}
Namely, the dynamics is diffusive, as verified numerically in Fig.~\ref{fig:random tumbling rate}.

\begin{figure}
\begin{center}
	\includegraphics[width=.5\textwidth]{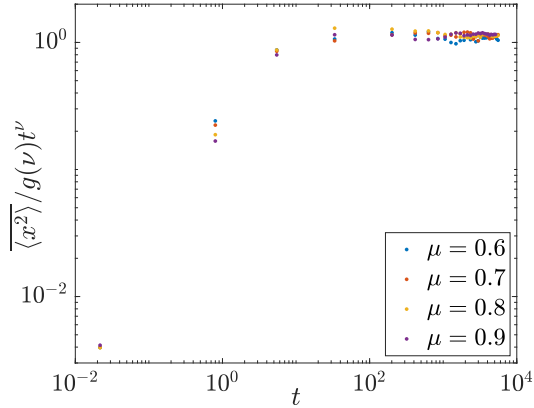} 
	\caption{The disorder-averaged spreading for a disordered tumbling rate (case II), rescaled by time according to Eq.~\eqref{eq:MFPT tumbling rate anomalous}. Here $\nu=2\mu/(1+\mu)$, and the rescaling factor $g(\nu)$ is chosen so that the curves asymptotically approach the same value. The dynamics of active particles is simulated with a random tumbling rate varying in space, drawn from the distribution $p(\alpha)\sim \alpha^{-(1+\mu)}$ for $\alpha\in (0,\infty)$, for 100 particles, $10^6$ time steps, 1000 realizations of disorder, and unit speed.}\label{fig:random tumbling rate divergent}
	\end{center}
\end{figure}

\noindent{\bf Case II: probability density with a diverging mean.}
To evaluate the expression for the MFPT~\eqref{eq:MFPT random tumbling rate}, we use the arguments presented in Case II of the speed disorder, taking $p(\alpha)\sim \alpha^{-(1+\mu)}$ for large $\alpha$ with $0<\mu<1$. Doing so, we find that the denominator scales as $L^{1/\mu}$ while the numerator as $L^{1+2/\mu}$. This leads to
\begin{equation}\label{eq:MFPT tumbling rate anomalous}
	\overline{\langle\tau\rangle} \underset{L\rightarrow\infty}{\propto} \frac{1}{2v^2}L^{1+1/\mu}\ ,
\end{equation}
a result  verified numerically in Fig.~\ref{fig:random tumbling rate divergent}.


\section{Many RTPs in a disordered potential}\label{sec:many particles}
As shown in the previous section, a RTP in a disordered
potential behaves as a random walker on a {\it random-forcing} energy landscape. This implies that the particle feels an effective potential whose depth grows as $\sqrt{L}$ with the system size $L$. In this section we consider the consequences of this fact for the many-body problem with a finite density of particles, for both interacting and non-interacting RTPs. Recall that for a single RTP in a disordered potential, there is a length scale $\ell^*$ (see Sec.~\ref{sec:MFPT potential}), below which the motion is diffusive and above which the motion is logarithmically slow. Accordingly, the discussion that follows is carried out separately for systems for which $L < \ell^*$, referred to as \emph{weak disorder}, and with $L > \ell^*$, referred to as \emph{strong disorder}. 

Most strikingly, we find that the presence of disorder promotes order in these systems. In the case of weak disorder, the two-point correlation function is shown, using numerics and a simplified theory, to decay linearly in space with an amplitude scaling linearly with the system size $L$. In the case of strong disorder with no interactions, as expected, the particles accumulate around a single minimum leading to a correlation function which decays over a finite distance. For the interacting case, it is well known that in the absence of disorder RTPs in one dimension only exhibit clusters of finite extent~\cite{Cates2015MIPS,tailleur2008statistical}. In contrast, here we find that the cluster size distribution is distributed as a power-law with a mean cluster size which scales as $\sqrt{L}$.

\subsection{Weak disorder}
\label{sec:many-body-weak-disord}

As explained above, we start by considering systems for which $L < \ell^*$. In this limit, the density profile of the system is expected to be approximately uniform with small fluctuations. Recalling that RTPs on a random potential are equivalent to Brownian particles on a random forcing energy landscape, we consider the simplified free energy functional 
\begin{equation}\label{eq:free}
	\mathcal{F} = \intop \textrm{d}x\, \left[\frac{K}{2}\left(\partial_x \phi\right)^2 + \frac{u}{2}\phi^2(x) + \phi(x) U(x)\right] \;.
\end{equation}
Here, the field $\phi(x)$ represents the density fluctuations from the mean value; $K$ accounts for interactions; $u$ contains both the leading order interaction and entropic contributions; while $U(x)$ is a random potential with the statistics of a random forcing energy landscape, 
\begin{equation}\label{eq:correlationsU}
\begin{cases}
 \overline{U(x)} = 0,\\ 
\overline{U(x)U(y)} = \frac{\sigma^2}{2}(x+y-|x-y|)\;.
\end{cases}
\end{equation}
Throughout this section, we use the convention $f_q=\frac{1}{\sqrt{L}}\int_0^L\textrm{d}x\, f(x)
e^{-iqx}$ for the Fourier transform of any function $f(x)$ defined on the interval $[0,L]$, e.g. with $\phi_q$ as the Fourier component $q$ of the density deviation field.

To characterize the effects of disorder, we focus on the disorder-averaged structure factor
\begin{equation}
  \label{eq:structure-factor}
  \overline{S(q)} = \overline{\langle\phi_q\phi_{-q}\rangle}\, ,
\end{equation}
with the overline, as before, denoting the average over disorder realizations and the brackets indicating  averages over the probability distribution governed by 
the Boltzmann weight from Eq.~\eqref{eq:free}.
Since this weight is Gaussian, one can easily calculate exactly the disorder-averaged structure factor:  The partition function $\mathcal{Z}$ for a given realization of the disorder is given by
\begin{equation}
	\ln\mathcal{Z}= \underset{q}{\sum}\, \left[\frac{1}{2u}\frac{U_q U_{-q}}{\left(1+\frac{K}{u}q^2\right)} - \ln\left(Kq^2+u\right)\right]\ ,
\end{equation}
where $U_q$ are the Fourier modes of the random potential. To compute the structure factor we first evaluate 
\begin{equation}
	\langle\phi_q\phi_{-q}\rangle_c = \frac{\delta}{\delta U_q}\frac{\delta}{\delta U_{-q}}\ln\mathcal{Z}\ ,
\end{equation}
with $\langle \phi_q\phi_{-q}\rangle_c = \langle\phi_q\phi_{-q}\rangle - \langle\phi_q\rangle \langle\phi_{-q}\rangle$  the connected correlation function. Then, noting that 
\begin{equation}
\langle\phi_q\rangle=\frac{1}{u}\frac{U_q}{\left(1+\frac{K}{u}q^2\right)}\ ,
\end{equation}
we arrive at the two-point correlation function  for a given realization of  disorder as
\begin{equation}\label{eq:rho_q rho_-q}
	S(q) = \frac{1}{u\left(1+\frac{K}{u}q^2\right)}\left[1+ \frac{1}{u}\frac{U_qU_{-q}}{\left(1+\frac{K}{u}q^2\right)}
\right]\ .
\end{equation}
After averaging over realizations of  disorder, we obtain
\begin{equation}
  \label{eq:structure-factor-Gaussian}
  \overline{S(q)} = \frac{1}{u\left(1+\frac{K}{u}q^2\right)}\left[1+ \frac{1}{u}\frac{\overline{U_qU_{-q}}}{\left(1+\frac{K}{u}q^2\right)}\right].
\end{equation}
Using Eq. \eqref{eq:correlationsU} to compute the correlation $\overline{U_qU_{-q}}$, we finally get for $q\neq0$
\begin{equation}
  \label{eq:structure-factor-Gaussian-f}
  \overline{S(q)} =\frac{1}{u\left(1+\frac{K}{u}q^2\right)}+\frac{2 \sigma^2}{q^2 u^2(1+{\frac{K}{u}q^2})^2}.
\end{equation}
To leading order in the $q\rightarrow0$ limit, one finds
$\overline{S(q)}\propto q^{-2}$ signalling that there are long-range correlations in the system. Note that this behavior is a consequence of the correlations in the potential, which are manifested even in the non-interacting case $K=0$, with $u$ accounting for purely entropic contributions. In real space, Eq. (\ref{eq:structure-factor-Gaussian-f}) gives the asymptotic behavior for large $r$ 
\begin{equation}
  \label{eq:structure-factor-real}
 \overline{S(r)} = \frac{1}{L}\intop_0^L \textrm{d}x\, \overline{\langle \phi(x)\phi(x+r) \rangle}\propto  L\left(1-A\frac{r}{L}\right),
\end{equation}
with $A$ an amplitude. The $\big(\frac{r}{L}\big)$ decay of Eq.~(\ref{eq:structure-factor-real}) shows that the correlations decay linearly with a scale proportional to the system size -- a result reminiscent of a phase separated system. This suggests that a macroscopic number of particles accumulate around the deepest part of the potential. 

Note that for $r=0$, we find that $\frac{1}{L}\int_0^L\textrm{d}x\, \overline{\phi^2(x)}\propto L$. This relation suggests a non-trivial scaling of the typical magnitude of the density fluctuations. To understand this scaling we note that the interaction term $\frac{u}{2}\phi^2(x)$ in the free energy accounts for repulsion between the particles and is balanced by the term $\phi(x)U(x)$, which tends to gather particles at the minimum of the potential. This minimum scales as $\sqrt{L}$, and so the typical magnitude of the density fluctuations in the dense phase scales in the same way.

\begin{figure}
\begin{center}
  \includegraphics[width=.5\textwidth]{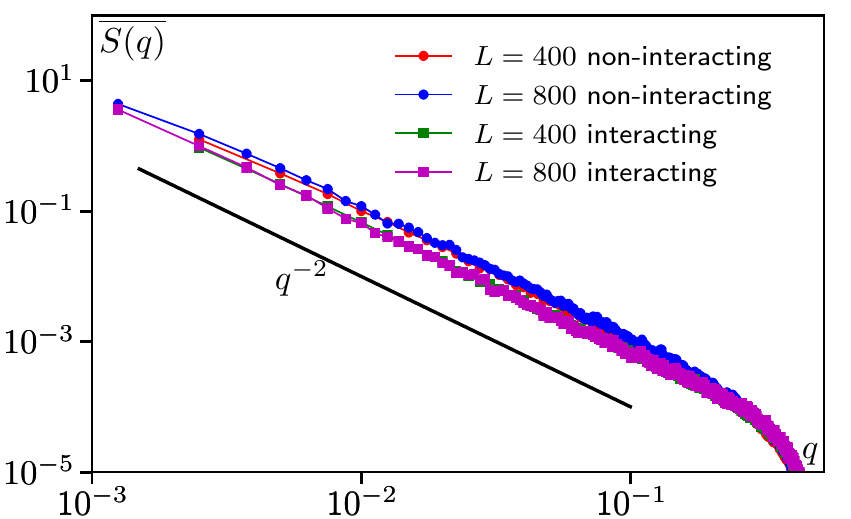} 
  \caption{Structure factor in the weak disorder case, with ratchets of height $V^*=0.58V^*_{\rm max}$. $V^*_{\rm max}$ is defined as the potential height which creates a slope $|\mu\partial_x V|=v$ which stops the particle.
  Interacting particles experience a short-range interaction of strength of $k=10$. The average density $\langle\rho\rangle=0.25$ is simulated on a ring of length $L=800$, with particles of unit speed, tumbling rate, and mobility.}\label{fig:weak-disord}
\end{center}
\end{figure}

We now compare the field-theoretic result with numerical simulations
of RTPs in the ratchet potential of Fig.~\ref{fig:ratchet
  potential}. We simulate, at the same average density of
$\rho_0=0.25$, both non-interacting particles and particles
interacting via a short-range pairwise harmonic repulsion
$U_r(\Delta)=\frac{k}{2}(1-\Delta)^2$ if $\Delta<1$ and $U_r=0$
otherwise, with $\Delta$ the inter-particle separation. The results
are presented in Fig.~\ref{fig:weak-disord}. As predicted by the field
theory, we find, for both interacting and non-interacting particles,
that the structure factor diverges as $q^{-2}$ at small $q$. For each
case, two different sizes are plotted in Fig.~\ref{fig:weak-disord} to
check that the scaling of the structure factor with $L$ is consistent
with the one given in Eq.~(\ref{eq:structure-factor-Gaussian-f}).

\subsection{Strong disorder}
\label{sec:many-body-strong-disord}

Here we consider the strong disorder limit where the system size obeys $L\gg\ell^*$, with $\ell^*$ the crossover length scale between the standard and ultra-slow diffusive regimes. Since numerical simulations proved prohibitively slow in this regime, we employ simple heuristic arguments building on the analogy between RTPs in a random potential and Brownian particles in a random forcing energy landscape. As we argue, unlike the weak disorder limit, the phenomenology is now very different between interacting and non-interacting particles.
 
\begin{figure}
\begin{center}
  \includegraphics[width=.5\textwidth]{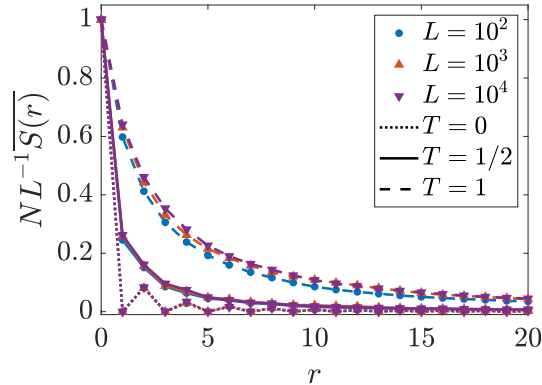}
  \caption{The disorder-averaged two-point function $\overline{S(r)} = L^{-1}\intop_0^x \textrm{d}x\,\overline{\langle\rho(x)\rho(x+r)\rangle}$ divided by the system size $L$ in real space, with periodic boundary conditions. To verify the scaling of Eq.~\eqref{eq:sacling nonint strong disorder}, the correlation function is rescaled by $L$ and plotted for different system sizes and temperatures of the random forcing equilibrium model. 
Since for non-interacting particles the overall density $\rho_0$ enters as a trivial $\rho_0^2 $ factor, we take $N$ such that $NL^{-1}\overline{S(r)}$ so that it is unity at $r=0$.
Note that the oscillations in the $T=0$ curves are due to degenerate minima.}\label{fig:c(r) nonint strong disorder}\label{fig:noninteracting}
\end{center}
\end{figure}

For the non interacting case, there is no bound on the maximal density at each point in space. 
At low enough temperatures $T\ll \sigma L^{\frac{1}{2}}$,
non-interacting particles can all collapse around the location of the global minimum of $U$ (as defined in Eq.~\eqref{eq:pot-U}), and the fluctuations are expected to be confined to a finite region around it. This leads to the expected behavior of the density-density correlation function
\begin{equation}\label{eq:sacling nonint strong disorder}
  \overline{{\cal S}(r)}=\frac{1}{L}\intop_0^L\textrm{d} x \,\overline{\langle \rho(x)\rho(x+r) \rangle} \sim  Lf(r),
\end{equation}
where $\rho(r)$ is the density at $r$, and $f(r)$ is a function that decays on a length scale independent of the length $L$.
As stated above, in the strong disorder regime, the convergence to the steady state proved too slow to obtain numerical results. We therefore use the analogy derived in section \ref{sec:potential disorder steady state} between active particles in a random potential and passive particles in a random forcing energy landscape. 
To this end, we use a potential defined on a lattice, such that the energy difference between adjacent sites is a random variable taking the values $\pm1$ (in arbitrary units) and therefore corresponds to a random force.
For non-interacting Brownian particles in a random-forcing energy landscape, the steady-state distribution is then given by the usual Boltzmann distribution $\rho_s(x)\propto e^{-U(x)/T}$, and can readily be 
evaluated.
(Recalling the discussion at the end of Sec.~\ref{sec:MFPT potential} we note that the strong disorder regime corresponds to low temperatures, and thus the Gaussian form emerging from Eq.~\eqref{eq:free} is no longer applicable in this  limit.) Evaluating the steady-state distribution numerically, we find that  as long as the temperature is low enough, the correlation functions  indeed behave as expected from Eq.~(\ref{eq:sacling nonint strong disorder}) (see Fig.~\ref{fig:c(r) nonint strong disorder}). This implies that essentially all the particles collapse near the minimal energy.

\begin{figure}
\begin{center}
  \includegraphics[width=\textwidth]{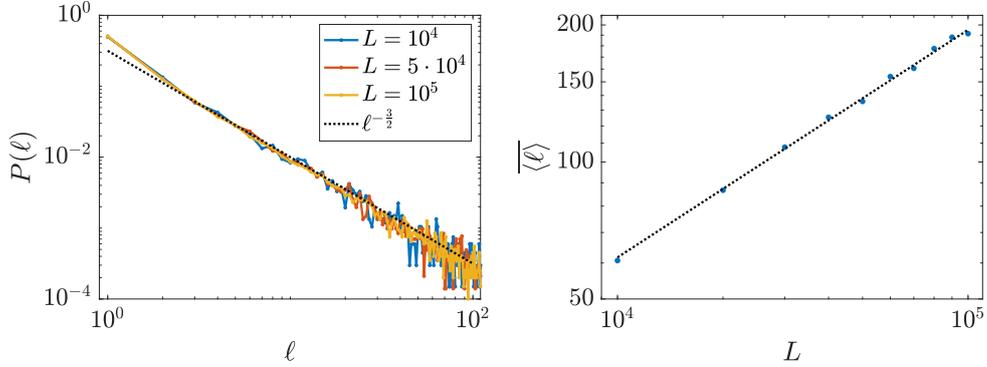}
  \caption{{\bf Left:} Distribution of cluster sizes for for interacting RTPs, with the same mean density as Fig.~\ref{fig:noninteracting} and different system sizes. The distribution is extracted from $10^3$ disorder realizations of the random forcing energy landscape. The curves are fitted with a probability distribution, scaling as $\ell^{-\frac{3}{2}}$. {\bf Right:} Average cluster size as a function of the system size, plotted along with the theoretical prediction of Eq.~\eqref{eq:l sqrt L}, $\overline{\langle\ell\rangle}\propto\sqrt{L}$.}\label{fig:cluster sizes}
\end{center}
\end{figure}
\begin{figure}
\begin{center}
  \includegraphics[width=.5\textwidth]{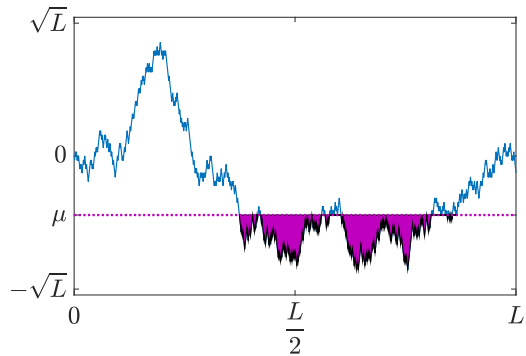}
  \caption{Particles with hard-core repulsion filling a realization of the random forcing energy landscape, at $T=0$. The particles occupy all energies up to the chemical potential, $\mu$, which controls the mean density, here set to $\langle\rho\rangle=0.4$.}\label{fig:chemical potential}
\end{center}
\end{figure}

We again employ the analogy to Brownian particles in a random forcing
potential to understand the case of interacting RTPs. For simplicity
we assume that the particles are on a lattice with hard-core
interactions. Such particles can be treated as non-interacting
Fermions with a chemical potential $\mu$ setting their overall
number. Clusters are defined as sequences of particles with no
vacancies and the distribution of cluster sizes is shown in
Fig.~\ref{fig:cluster sizes} (left). (Since the overall energy scale,
set by the chemical potential, is much larger than the effective
temperature we consider the zero temperature limit of this model.)
Interestingly, the domain size distribution, $P(\ell)$ behaves as a
power-law
\begin{equation}
\label{eq:ellstat}
	P(\ell) \sim \frac{1}{\ell^{3/2}} \;,
\end{equation}
implying that the average domain size in the system grows as
\begin{equation}\label{eq:l sqrt L}
	\langle \ell \rangle \sim \sqrt{L}\;,
\end{equation}
as verified in Fig.~\ref{fig:cluster sizes} (right). 
The origin of the power-law distribution of cluster sizes can be understood by considering Fig.~\ref{fig:chemical potential}, in which the chemical potential, controlling the filled locations on the lattice, is marked explicitly for a given realization of the disorder. It is clear that the size of a cluster is dictated by the statistics of first return of a random walk, starting and ending at the chemical potential, which is indeed governed by Eq.~\eqref{eq:ellstat}. 

\begin{figure}
\begin{center}
  \includegraphics[width=.5\textwidth]{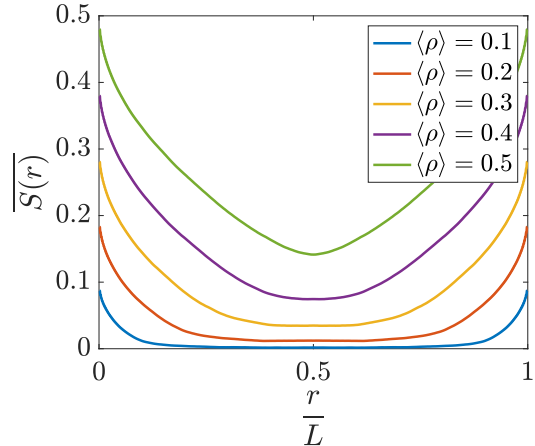}
  \caption{The disorder-averaged two-point function $\overline{S(r)} = L^{-1}\intop_0^x \textrm{d}x\,\overline{\langle\rho(x)\rho(x+r)\rangle}$ in real space, with periodic boundary conditions. The curves correspond to different average densities, $\langle\rho\rangle$.}\label{fig:c(r)}
\end{center}
\end{figure}

We end this section by noting that it is straightforward to numerically obtain the two-point correlation function, $\overline{{\cal S}(r)}$ (see Fig.~\ref{fig:c(r)}). Simple theoretical arguments using  the self-similarity of the Brownian motion show that for a fixed particle density, the two-point correlation function is a function of $\big(\frac{r}{L}\big)$. However, we have not been able to obtain explicit analytical expressions for this functional form. The problem is related to statistics of extrema of random walks (see for example~\cite{perret2013near,majumdar2014gap} where the problem is properly defined but still unsolved) and is beyond the scope of this manuscript.

\section{Summary}\label{sec:summary}

In this paper we studied active particles in disordered
one-dimensional environments. Considering the effects of
three types of disorder (in the external potential, the speed, and the
tumbling rate), we derived the steady-state distributions and
dynamical properties for a single run-and-tumble particle. In the case
of potential disorder, we also consider the many-body case with either
interacting or non-interacting particles.

In the single-particle problem, the most striking manifestation of
disorder was obtained for random potentials: The active particles were
shown to exhibit ultra-slow Sinai diffusion at long times, a behavior
analogous to passive random walkers on a {\it random forcing} energy
landscape. In the paper, we considered run-and-tumble particles with
no translational diffusion and were therefore restricted to potentials
with limited slope. If translational diffusion is allowed, the
particles can hop across steep local barriers, as recently analyzed in
Ref.~\cite{woillez2019activated}. Using these results, it is easy to
see that a local forcing is still present in this case when the
barrier lacks an inversion symmetry. Therefore, the conclusion drawn
from the case analyzed in the bulk of this paper remains
unchanged. Similarly, any 1d active particle model where ratchet
currents are generated should exhibit the same phenomenology.

For disordered speed and tumbling rate, we have shown that, as long as
the mean tumbling rate and the mean inverse speed are finite, the
active particles exhibit ordinary diffusion. If, on the other hand,
these averages diverge, the diffusion of the particles becomes
anomalous.

In the many-body problem, we also found that a random potential leads
to striking effects. The disorder-averaged structure factor was
evaluated in the weak disorder regime using a field theory and was
shown to diverge as $q^{-2}$ at small wave vectors. In the strong
disorder regime, the phenomenology is different. Using the passive
random forcing model, we found that while non-interacting particles
all aggregate at a particular locus, interacting particles form much
wider clusters, with the average cluster width scaling as the square
root of the system size.

It should be noted that while the paper deals only with one
dimensional disordered systems, it offers a hint on the dynamics in
higher dimensions. As the arguments presented above are rather
general, we expect that our results can be extrapolated to higher
dimensions. In two-dimensional random potential models, for example,
circulation currents would appear~\cite{chepizhko2013diffusion} due to
the effective random forces induced by the potential. In this case,
the disorder-averaged spreading should follow the random forcing
diffusive scaling, $\overline{\langle x^2 \rangle} \sim
t/\ln{(t)}$. Furthermore, in the case of a disordered speed or
tumbling rate, we expect results similar to those found for one
dimension to persist in higher dimensions.

\noindent {\it Acknowledgments}: YBD, EW \& YK acknowledge support from the Israel Science Foundation. YBD, EW, YK \& MK were also supported by an NSF5-BSF grant (DMR-170828008).

\bibliography{Disorder}

\newpage

\appendix
\renewcommand{\theequation}{\thesection.\arabic{equation}}
\section{Computation of the mean first passage time}
\setcounter{equation}{0}

The mean first passage time (MFPT) $\langle\tau\rangle$ of particles absorbing at a distance $L$ from the origin can be computed exactly.
In Section~\ref{app:computation of MFPT random potential} the calculation is done for the random potential, and for both the cases of random speed and tumbling rate in Section~\ref{app:computation of MFPT random speed and tumbling rate}. For completeness, the derivation is presented without assuming much background. 

\subsection{Random potential}
\label{app:computation of MFPT random potential}

In this section, the MFPT is computed for a quenched-disordered potential, as defined in Section~\ref{sec:potential disorder steady state}.
We introduce the MFPT $\tau_{+}(x)$ (resp. $\tau_{-}(x)$) of a particle initially located at position $x$ and moving in the right (resp. left) direction.
As expected, the long time scale behavior of the two expressions, which is of interest, will be the same, resulting in a single expression for the MFPT irrespective of the initial condition. 
 
The calculation is done by employing the backward Fokker-Planck equation~\cite{gardiner1986handbook}. To this end, we consider the backward evolution of the probability density $P_\pm(x',t;x,0)$ of particles reaching $X=x'$ at time $t$, initially starting at $X=x$ moving respectively to the right or to the left~\cite{malakar2018steady}
\begin{align}\label{eq:FPback}
\partial_{t}P_{+}\left(x',t;x,0\right)=&\left[v-\mu\left(\partial_{x}V\right)\right]\partial_{x}P_{+}\left(x',t;x,0\right)-\frac{\alpha}{2}\left(P_{+}\left(x',t;x,0\right)-P_{-}\left(x',t;x,0\right)\right)\ ,\nonumber\\
\partial_{t}P_{-}\left(x',t;x,0\right)=&\left[-v-\mu\left(\partial_{x}V\right)\right]\partial_{x}P_{-}\left(x',t;x,0\right)+\frac{\alpha}{2}\left(P_{+}\left(x',t;x,0\right)-P_{+}\left(x',t;x,0\right)\right)\ .
\end{align}
These equations are solved with the absorbing boundary conditions, $P_+(x',t;L,0)=P_-(x',t;-L,0)=0$.
For compactness, time can be rescaled using the inverse tumbling rate and length can be rescaled by the factor $\frac{v\alpha}{2}$. In these non-dimensional units, we denote $\left(-\mu\partial_xV\right)$ by $\varphi(x)$. Equations~(\ref{eq:FPback}) can then be written in the dimensionless form
\begin{align}
\partial_{t}P_{+}\left(x',t;x,0\right)=&\left(1+\varphi\left(x\right)\right)\partial_{x}P_{+}\left(x',t;x,0\right)-\left(P_{+}\left(x',t;x,0\right)-P_{-}\left(x',t;x,0\right)\right)\nonumber\ ,\\
\partial_{t}P_{-}\left(x',t;x,0\right)=&\left(-1+\varphi\left(x\right)\right)\partial_{x}P_{-}\left(x',t;x,0\right)+\left(P_{+}\left(x',t;x,0\right)-P_{-}\left(x',t;x,0\right)\right)\ .\label{eq:Backward FP dimensionless}
\end{align}

The probability that a particle
is not absorbed in a time interval $t$,
$G\left(x,t\right)$, conditioned that it was initially positioned
at $X=x$, is given by the spatial integral over its final position~\cite{gardiner1986handbook}
\begin{equation}
G_{\pm}\left(x,t\right)\equiv\intop_{-L}^{L}\textrm{d}x'\,P_{\pm}\left(x',t;x,0\right)\ ,
\end{equation}
with the initial condition $G_{\pm}\left(x,0\right)=1$ for $-L<x<L$. Integrating
Eq. (\ref{eq:Backward FP dimensionless}) over $x'$ shows that  $G_{\pm}\left(x,t\right)$
satisfy the backward Fokker-Planck equations
\begin{align}
    \partial_{t}G_{+}\left(x,t\right)=&\left(1+\varphi\left(x\right)\right)\partial_{x}G_{+}\left(x,t\right)-\left(G_{+}\left(x,t\right)-G_{-}\left(x,t\right)\right)\ ,\nonumber\\
    \partial_{t}G_{-}\left(x,t\right)=&\left(-1+\varphi\left(x\right)\right)\partial_{x}G_{-}\left(x,t\right)+\left(G_{+}\left(x,t\right)-G_{-}\left(x,t\right)\right)\ . \label{eq:BFP G}
\end{align}
The functions $G_\pm\left(x,t\right)$ give the probability that absorption of the particle happens after time $t$, and are related to the probability densities $\rho_{\pm}(\tau)$ of the first passage time through
\begin{equation}
G_\pm\left(x,t\right)=\intop_t^{\infty}\rm{d}\tau\,\rho_{\pm}(\tau)\;,
\end{equation}
which gives after differentiation $\rho_{\pm}(t)=-\partial_{t}G\left(x,t\right)$.
The latter relation can be eventually used to compute the MFPT through
\begin{align}\label{eq:end of MFPT general derivation}
\tau_{\pm}\left(x\right)= & -\intop_{0}^{\infty}{\rm d}t\,t\partial_{t}G_{\pm}\left(x,t\right)\nonumber \\
= & \intop_{0}^{\infty}{\rm d}t\,G_{\pm}\left(x,t\right)\ .
\end{align}
Thus, integrating Eq.~(\ref{eq:BFP G}) over time, the MFPTs
$\tau_{\pm}\left(x\right)$ are shown to obey the following backward stationary
Fokker-Planck equations
\begin{align}
    \left(1+\varphi\left(x\right)\right)\partial_{x}\tau_{+}\left(x\right)-\left(\tau_{+}\left(x\right)-\tau_{-}\left(x\right)\right)&=-1\ ,\nonumber\\
    \left(-1+\varphi\left(x\right)\right)\partial_{x}\tau_{-}\left(x\right)+\left(\tau_{+}\left(x\right)-\tau_{-}\left(x\right)\right)&=-1\ .\label{eq:BFP t-1}
\end{align}

Equation~(\ref{eq:BFP t-1})  can be solved using the boundary conditions $\tau_+(L)=\tau_-(-L)=0$, which mean that a particle starting on the boundary with its velocity pointing outwards is immediately absorbed. We choose to skip the lengthy calculations and give directly the general expressions for $\tau_{\pm}(x)$. As we are interested in the time required for a particle to travel a distance of $L$, we only consider particles starting at $x=0$, with equal probability of moving to the left or to the right. For convenience we define the function
\begin{equation}\label{eq:psi def}
	\psi(x) =  \exp\left(\,\intop_{-L}^{x}\textrm{d}y\,\frac{\varphi(y)}{1-\varphi^2(y)}\right)\ .
\end{equation}
The mean first passage time $\langle\tau\rangle$ is then computed as
\begin{subequations}\label{eq:tplus appendix}
\begin{align}
\langle\tau\rangle =& \frac{1}{2}\bigg[\tau_{+}\left(0\right)+\tau_{-}\left(0\right)\bigg] \\
= & \left\{ \psi\left(0\right)\left[\intop_{-L}^{L}\textrm{d}z\,\frac{\psi\left(z\right)}{1-\varphi^{2}\left(z\right)}\right]\right.\\
 & -\psi\left(L\right)\left[\intop_{-L}^{L}\frac{\textrm{d}y}{\psi\left(y\right)}\frac{2}{1-\varphi^{2}\left(y\right)}\right]\left[\intop_{-L}^{0}\textrm{d}z\,\frac{\psi\left(z\right)}{1-\varphi^{2}\left(z\right)}\right]\\
 & +\frac{1}{2}\psi\left(0\right)\left[\intop_{-L}^{L}\textrm{d}y\,\frac{{\rm sgn}(y)}{\psi\left(y\right)}\frac{2}{1-\varphi^{2}\left(y\right)}\right]\left[\intop_{-L}^{L}\textrm{d}z\,\frac{\psi\left(z\right)}{1-\varphi^{2}\left(z\right)}\right]\\
 & -\frac{1}{2}\psi\left(0\right)\left[\intop_{-L}^{L}\textrm{d}y\,\frac{{\rm sgn}(y)}{1-\varphi^{2}\left(y\right)}\left(\varphi\left(y\right)-\intop_{-L}^{y}\textrm{d}z\,\frac{\psi\left(z\right)}{\psi\left(y\right)}\frac{2}{1-\varphi^{2}\left(z\right)}\right)\right]\\
& +\frac{1}{2}\psi\left(0\right)\psi\left(L\right)\left[\intop_{-L}^{L}\textrm{d}y\,\frac{{\rm sgn}(y)}{1-\varphi^{2}\left(y\right)}\left(\varphi\left(y\right)-\intop_{-L}^{y}\textrm{d}z\,\frac{\psi\left(z\right)}{\psi\left(y\right)}\frac{2}{1-\varphi^{2}\left(z\right)}\right)\right]\\
 & +\psi\left(0\right)\psi\left(L\right)\left[\intop_{0}^{L}\frac{\textrm{d}z}{\psi\left(w\right)}\frac{2}{1-\varphi^{2}\left(w\right)}\right]\left[\intop_{-L}^{0}\frac{\textrm{d}y}{1-\varphi^{2}\left(y\right)}\left(\varphi\left(y\right)-\intop_{-L}^{y}\textrm{d}z\,\frac{\psi\left(z\right)}{\psi\left(y\right)}\frac{2}{1-\varphi^{2}\left(z\right)}\right)\right]\label{eq:MFPT appendix 1st}\\
 & \left.-\psi\left(0\right)\psi\left(L\right)\left[\intop_{-L}^{0}\frac{\textrm{d}z}{\psi\left(w\right)}\frac{2}{1-\varphi^{2}\left(w\right)}\right]\left[\intop_{0}^{L}\frac{\textrm{d}y}{1-\varphi^{2}\left(y\right)}\left(\varphi\left(y\right)-\intop_{-L}^{y}\textrm{d}z\,\frac{\psi\left(z\right)}{\psi\left(y\right)}\frac{2}{1-\varphi^{2}\left(z\right)}\right)\right]\right\} \label{eq:MFPT appendix 2nd}\\
& \times\left\{\psi\left(0\right)\left[\psi\left(L\right)\left[\intop_{-L}^{L}\frac{du}{\psi\left(u\right)} \frac{2}{1-\varphi^{2}\left(u\right)}+1\right]+1\right]\right\}^{-1}\,.
\end{align}
\end{subequations}
Note that the seemingly asymmetric expression of $\langle\tau\rangle$ is due to the definition of $\psi(x)$ in Eq.~\eqref{eq:psi def}.

\subsection{Random speed and tumbling rate}
\label{app:computation of MFPT random speed and tumbling rate}

To compute the MFPT for models of random speed or tumbling rate, we follow the method detailed in the random potential case~\ref{app:computation of MFPT random potential}. As $v(x)$ or $\alpha(x)$ are no longer constants, the computation is done with dimensional quantities, leading to the following set of equations
\begin{align}
	v(x)\partial_{x}\tau_{+}\left(x\right)-\frac{\alpha(x)}{2}\bigg[\tau_{+}\left(x\right)-\tau_{-}\left(x\right)\bigg]&=-1\ ,\nonumber\\
	-v(x)\partial_{x}\tau_{-}\left(x\right)+\frac{\alpha(x)}{2}\bigg[\tau_{+}\left(x\right)-\tau_{-}\left(x\right)\bigg]&=-1\ .
\end{align}

Solving these equations, the MFPT reads
\begin{subequations}
\begin{align}
	\langle\tau\rangle =& \frac{1}{2}\bigg[\tau_{+}\left(0\right)+\tau_{-}\left(0\right)\bigg] \\
	= & \Bigg[\ \intop_{-L}^{L}\frac{\textrm{d}x}{v\left(x\right)} +\intop_{-L}^{0}\textrm{d}x\,\frac{\alpha\left(x\right)}{v\left(x\right)}\intop_{x}^{L}\frac{\textrm{d}y}{v\left(y\right)}+\intop_{0}^{L}\textrm{d}x\,\frac{\alpha\left(x\right)}{v\left(x\right)}\intop_{-L}^{x}\frac{\textrm{d}y}{v\left(y\right)} \\
	& +\intop_{-L}^{0}\textrm{d}x\,\frac{\alpha\left(x\right)}{v\left(x\right)}\intop_{0}^{L}\textrm{d}y\,\frac{\alpha\left(y\right)}{v\left(y\right)}\intop_{x}^{y}\frac{\textrm{d}z}{v\left(z\right)}\Bigg]\label{eq:dominating MFPT speed and tumbling rate appendix}\\
	& \times\left[2+\intop_{-L}^{L}\textrm{d}z\,\frac{\alpha\left(w\right)}{v\left(w\right)}\right]^{-1}\ .
\end{align}
\end{subequations}
As there are no exponential terms in this result, the dominating term
in the large length scale limit is found by analyzing the large $L$
behavior and is that of Eq.~\eqref{eq:dominating MFPT speed and
  tumbling rate appendix}.  This result is used in the main text for
random speed~\ref{sec:velocity MFPT} and random tumbling
rate~\ref{sec:tumbling rate MFPT}.

\end{document}